\documentclass[twocolumn, preprintnumbers,amsmath,amssymb]{revtex4}
\usepackage{txfonts}
\usepackage{amssymb}
\usepackage{setspace}
\usepackage{graphicx}
\usepackage{subfigure}

\begin{document}

\title{Spin Dephasing in Drift-Dominated Semiconductor Spintronics Devices}
\author{Biqin Huang}
\altaffiliation{bqhuang@udel.edu}
\author{Ian Appelbaum}
\affiliation{ Electrical and Computer Engineering Department,
University of Delaware, Newark, Delaware, 19716}

\begin{abstract}
A spin transport model is employed to study the effects of spin dephasing induced by diffusion-driven transit-time uncertainty through semiconductor spintronic devices where drift is the dominant transport mechanism. It is found that in the ohmic regime, dephasing is independent of transit length, and determined primarily by voltage drop across the spin transport region. The effects of voltage and temperature predicted by the model are compared to experimental results from a 350-micron-thick silicon spin-transport device using derived mathematical expressions of spin dephasing.
\end{abstract}

\maketitle
\newpage

\section{Introduction}

Employing electron spin in semiconductor devices could potentially overcome scaling issues with modern charge-based electronics, providing more power efficiency, higher speed, and greater functionality\cite{WOLF2001, WOLF2006, AWSCHALOM2007, GREGG2002, ZUTIC2004, DERYNATURE, SLOVACA}. In many spintronic device designs, spin precession during transport from injector to detector plays a primary role in determining device output characteristics. This precession is caused by a torque exerted either by effective\cite{DATTADAS} or real\cite{SPINFETTHRY, 37PERCENT} magnetic fields oriented perpendicular to the spin direction. 

Because of magnetic field inhomogeneities or transit time uncertainty, transported spins will arrive at the spin detector with a distribution of precession angles. When the angular distribution width approaches 2$\pi$, the contributions of spin-polarized electrons to the detector output is reduced by partial signal cancellation of oppositely-oriented spins, in a process called ``dephasing'', or ``Hanle effect''\cite{MONZON, JOHNSON1985,JOHNSON1988}. 

Dephasing in spin-transport devices (especially when due to transit-time uncertainty caused by spatial diffusion or pathlength variation) cannot be reversed, as it can be with spin-echo techniques in electron spin resonance (where transport plays no role)\cite{HAHN}. Therefore, spin dephasing is a physical process of fundamental importance to semiconductor spintronic devices. Here, we present analysis of a spin drift-diffusion model to quantify the effects of dephasing in a regime where charge transport is dominated by drift, as in recent experiments\cite{LOU, APPELBAUMNATURE, HUANG350}. We examine the apparent independence of relative dephasing (precession fringe visibility in magnetic field spectroscopy) to injector-detector distance (transit length) and quantify the effects of voltage bias across the transport region. These conclusions are bolstered through comparison to experiment, using silicon spin-transport devices\cite{APPELBAUMNATURE}. In addition, the effects of temperature on spin dephasing are discussed.

\section{Model}

Since spin detection is invariably a projection of the electron spin on a fixed measurement axis, the change in output of a spintronics device employing precession will be determined by $\cos{\theta}$ (to first order), where $\theta=\omega\cdot\tau$ is spin angle, and $\omega$ is spin precession frequency and $\tau$ is transit time from injector to detector over transit length $L$. Clearly, transit-time uncertainty $\Delta \tau$ gives rise to precession angle uncertainty $\Delta\theta=\omega\cdot\Delta\tau$. Increasing precession frequency ($\omega=g\mu_B B/ \hbar$, where $g$ is electron spin g-factor, $\mu_B$ is the Bohr magneton, $\hbar$ is the reduced Planck's constant, and $B$ is perpendicular magnetic field) by increasing the magnetic field therefore increases the angular uncertainty $\Delta \theta$ and causes dephasing. A complete model for dephasing during transport from injector to detector is therefore needed.

Electron transport in a semiconductor is well-described by the drift-diffusion equation. To model spin transport accurately, finite spin-lifetime must be included within the relaxation-time approximation:

\begin{equation}\label{eqn:spindriftdiffusion}
\frac{\partial s}{\partial t}=D\frac{\partial^2 s}{\partial
x^2}-v\frac{\partial s}{\partial x}-\frac{s}{\tau_{sf}},
\end{equation}

\noindent where $s$ is the spin density, $\tau_{sf}$ represents the spin relaxation time, $D$ is the diffusion coefficient for electrons, and $v$ is the electron drift velocity. In the ``ohmic'' regime, $v=\mu E$, where $\mu$ is mobility and $E$ is electric field. Eq. \ref{eqn:spindriftdiffusion} governs the evolution of spatial distributions of electron spin, and can be solved easily for initial conditions (at $t=0$) of an ensemble of spins which are all spin-polarized in the same direction at the injector (at $x=0$). In other words, we find the Green's function describing the evolution of a Dirac delta:

\begin{equation}\label{eqn:spindriftanddiffusion}
 s(x,t)=\frac{1}{2\sqrt{\pi Dt}}e^{-\frac{(x-vt)^2}{4Dt}}e^{-\frac{t}{\tau_{sf}}}.
\end{equation}

The most salient features of this gaussian solution for our subsequent analysis are that the center of the distribution travels with velocity $v$ toward the detector, and that the spatial distribution width increases in time as $\sqrt{D t}$. 

Holding $x=L$, this solution describes the distribution of arrival times $t$ at the detector. Therefore, the precession-induced change in total device output signal ($\Delta I_c$) comprised from all electrons arriving with different transit times $0<t<\infty$, each contributing $\cos{\theta}$ due to spin precession, is 

\begin{equation}\label{eqn:precession2}
\Delta{I_c} \propto \int_0^\infty\left[ \frac{1}{2\sqrt{\pi
Dt}}e^{-\frac{(L-vt)^2}{4Dt}}e^{-\frac{t}{\tau_{sf}}}\right] \cos{\omega t}dt.
\end{equation}

This integral expression can be easily evaluated to compare to empirical observations. A typical experiment consists of a spectroscopy where perpendicular magnetic field strength $B$ is varied; spin precession oscillations (where extrema correspond to average precession angles in integer multiples of $\pi$) are suppressed in larger magnetic fields due to a larger precession-angle uncertainty. This suppression due to dephasing, or ``Hanle effect'', can be seen in results evaluated from Eq. \ref{eqn:precession2} shown in Fig. \ref{fig:fig2}(a). More relative dephasing increases spin decoherence at the detector and reduces the number of oscillations seen.

\section{Dephasing Analysis}

\begin{figure}
  \centering
  \includegraphics[scale=0.775]{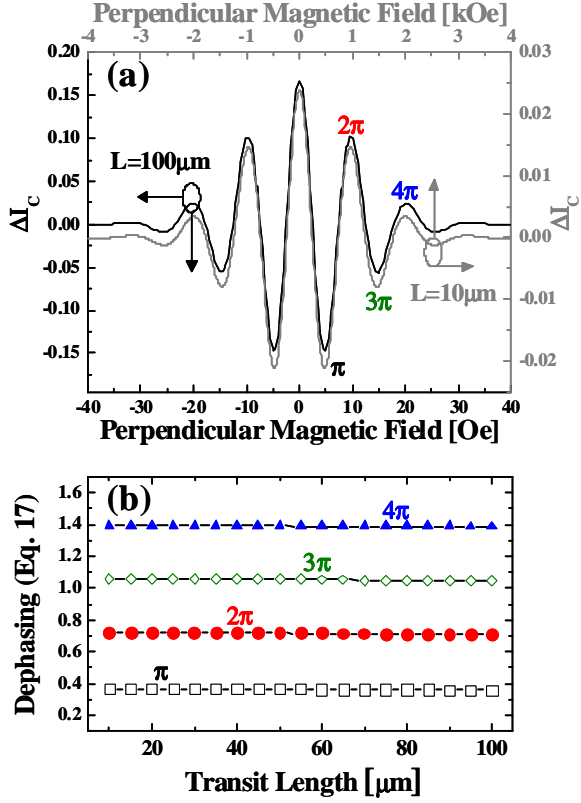}
  \caption{ \label{fig:fig2} (a) Comparison of the spin precession signal output according to Eq. \ref{eqn:precession2} for devices with  $L=100 \mu m$ (black) and $10 \mu m$ (light grey) transport layer thickness, showing identical dephasing. The voltage bias is 2V, mobility is 1400 cm$^2$/Vs, diffusion coefficient is 36 cm$^2$/s, and spin lifetime is 100ns for both simulations. Several oscillation extrema are labeled by the average precession angle $\pi$, $2\pi$, $3\pi$, $4\pi$. (b) The independence of dephasing to transport length regardless of oscillation order is shown using Eq. \ref{eqn:dephasing0} calculated for $10\mu$ m$<L<100\mu$ m. }
\end{figure}

Because the integrand in Eq. \ref{eqn:precession2} is dominated by the gaussian term $e^{-\frac{(L-vt)^2}{4Dt}}$, the exponentially decaying part $e^{-\frac{t}{\tau_{sf}}}$ can be ignored if the spin lifetime $\tau_{sf}$ is suitably long. Then we have

\begin{equation}\label{eqn:temp101}
\Delta{I_c} \sim \int_0^\infty\frac{1}{2\sqrt{\pi Dt}}e^{-\frac{(L-vt)^2}{4Dt}}\cos{\omega t}dt.
\end{equation}

Making the variable substitution $\theta \equiv \omega t$ converts Eq. \ref{eqn:temp101} to 

\begin{equation}\label{eqn:temp102}
\Delta I_c\sim \int_0^\infty\frac{1}{2\sqrt{\pi D\theta\omega}}e^{-\frac{(L\omega^2-v\theta\omega)^2}{4D\theta}}\cos{\theta}d\theta.
\end{equation}

Application of the following transformation

\begin{eqnarray}\label{eqn:transformation}
L & \longrightarrow & AL \nonumber \\
v &\longrightarrow& \frac{v}{A}\nonumber \\
\omega & \longrightarrow & \frac{\omega}{A^2} 
\end{eqnarray}

\noindent (where $A$ is a transit-length scaling factor) converts Eq. \ref{eqn:temp102} to

\begin{eqnarray}\label{eqn:temp103}
\Delta{I_c} &\sim &\int_0^\infty\frac{1}{2\sqrt{\pi
D\theta\frac{\omega}{A^2}}}e^{-\frac{\frac{\omega}{A^2}(AL-\frac{v}{A}\frac{\theta A^2}{\omega})^2}{4D\theta}}\cos{\theta}d\theta\nonumber\\
& = &A\int_0^\infty\frac{1}{2\sqrt{\pi D\theta\omega}}e^{-\frac{(L\omega^2-v\theta\omega)^2}{4D\theta}}\cos{\theta}d\theta.
\end{eqnarray}

Other than an overall multiplicative factor of $A$, Eq. \ref{eqn:temp103} is identical to Eq. \ref{eqn:temp102}; the spin precession model is therefore virtually invariant to this transformation. In terms of device physics, ohmic transport ($v=\mu V/L$, where $V$ is the voltage bias on the transport layer) automatically accounts for decreasing the drift velocity by a factor of $A$ if the length $L$ increases by the same proportion at constant voltage; the first two elements of the transformation are therefore satisfied. The third element of the transformation in Eq. \ref{eqn:transformation} is accounted for by applying a quadratically weaker magnetic field. The experimental outcome of this result is that in measurements of different devices, we can expect the same number of precession oscillations despite variation in transport lengths, assuming the applied voltage $V$ is identical.

Again, the preceding assumes that the exponential spin-decay term is negligible. This is true when the spin lifetime is much longer than the time it takes the gaussian spin distribution of width $d$ to enter into the detector (transit time uncertainty, $\Delta \tau=\frac{d}{v}$). Since for drift-dominated operation 

\begin{equation}\label{eq:diffbroaden}
d=\sqrt{D\tau}=\sqrt{D\frac{L}{v}}, 
\end{equation}

\noindent we have

\begin{equation}
\Delta{\tau}=\sqrt{\frac{DL}{v^3}}.
\end{equation}

If $\tau_{sf}\gg \Delta{\tau}$, then finite spin-lifetime can be ignored and variations in $L$ make negligible difference to the evaluation of Eq. \ref{eqn:precession2}. For characteristic values $D=100$cm$^2$/s, $v=10^6$cm/s, and $\tau_{sf}=100$ns, this condition is satisfied for $L\ll 100$cm, which is certain to hold for semiconductor devices that are typically many orders of magnitude smaller.

Fig. \ref{fig:fig2}(a) shows simulation results for different device transport layer thicknesses $L$. The black and light grey curve correspond to $L_1 =100 \mu $m and $L_2=AL_1=10 \mu$ m, respectively. The simulation is performed using $\mu=1400$cm$^2$/Vs, $D=36$ cm$^2$/s and voltage bias $V=2$ V. The spin lifetime in both simulations is 100ns. For the sake of comparison, the $y$ axes are slightly shifted relative to each other. It is very clear that the spectroscopy shape and precession oscillation extrema number is exactly the same, despite the difference in signal magnitude, which is roughly equal to $A$ (a factor of 10). 

Another way to show the dephasing invariance to transit length is to calculate the relative uncertainty in the distribution of final precession angle

\begin{equation}\label{eqn:length0}
\frac{\Delta\theta}{\theta}=\frac{\omega \cdot \Delta \tau}{\omega\tau}.
\end{equation}

\begin{figure}
  \centering
  \includegraphics[scale=0.6]{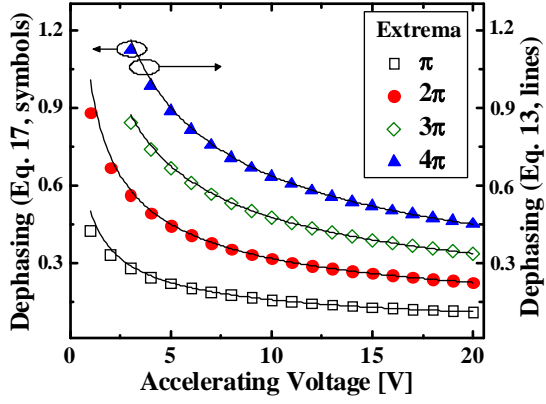}
  \caption{ \label{fig:fig3}Dephasing (defined by Eq. \ref{eqn:dephasing0}) for the precession oscillation extrema calculated with Eq. \ref{eqn:precession2} as a function of applied voltage across a 350 $\mu$m transport region with mobility $\mu=1400$cm$^2$/Vs (symbols). Lines are analytically given by Eq. \ref{eqn:length} with corresponding values of $\theta$ indicated by the legend. Note exceptional agreement between these two complementary means of determining dephasing.
 }
\end{figure}

If transport is dominated by drift in the applied electric field\cite{BIQINJAP}, the transit time is given by

\begin{equation}\label{eqn:transittime}
\tau=\frac{L}{v}=\frac{L}{\mu E}=\frac{L}{\mu \frac{V}{L}}=L^2/(\mu V).
\end{equation}

\noindent (Note the quadratic dependence of transit time on transport length at fixed voltage.) Applying ohmic transport $v=\mu V/L$ to Eq. \ref{eq:diffbroaden} gives

\begin{equation}
d=L\sqrt{D/(\mu V)}.
\end{equation}

Since the width of the transit time distribution $\Delta \tau$ is $d/v$, the uncertainty in the distribution of final precession angle $\theta$ at the detector is given by application of Eq. \ref{eqn:length0}:

\begin{equation}\label{eqn:length}
\Delta\theta=\theta\sqrt{D/(\mu V)}.
\end{equation}

This result (valid at precession extrema when the average spin direction is either parallel or antiparallel to the measurement axis) is independent of the transit length $L$, so that we can expect the same amount of dephasing at the same voltage bias through the transport layer, regardless of the distance from injector to detector for any fixed precession angle (assuming ohmic behavior, $v=\mu E$, where $E$ is internal electric field). Although Eq. \ref{eqn:length} is independent of transit length $L$, it is clearly dependent on voltage bias $V$. Specifically, it predicts that dephasing is suppressed with an increase in $V$, which is intuitively expected since diffusion will play a smaller role as drift becomes stronger in a larger electric field.

\begin{figure}
  \centering
  \includegraphics[scale=0.45]{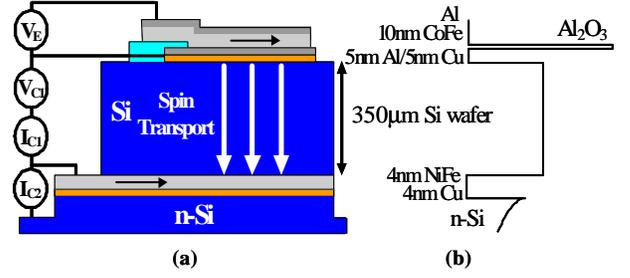}
  \caption{ \label{fig:fig4} (a) Side-view and (b) schematic conduction band diagram of our 350-micron-thick undoped single-crystal silicon spin transport device used to compare experimental results to the dephasing model presented.  
 }
\end{figure}

To further study the dephasing effect, it is necessary to establish a proper quantitative empirical standard for comparison to experimental measurements not dependent on data fitting using evaluation of Eq. \ref{eqn:precession2}. If drift dominates and $L/v\gg\Delta\tau$, the factor proportional to $1/\sqrt{t}$ in Eq. \ref{eqn:precession2} can be approximated as a constant. Then, (assuming again that $\tau_{sf}\gg\Delta\tau$) the distribution of precession angles is gaussian:

\begin{equation}
\frac{1}{\sqrt{2\pi}\Delta\theta}e^{-\frac{\theta^2}{2\Delta\theta^2}}
\end{equation}

\noindent so we have

\begin{equation}
\Delta I_c \sim \int^{+\infty}_{-\infty} \frac{1}{\sqrt{2\pi}\Delta\theta}e^{-\frac{\theta^2}{2\Delta\theta^2}} \cos{\theta} d \theta = e^{-\frac{\Delta\theta^2}{2}}.
\end{equation}

We can use this analytic result to find the angular distribution width from $\Delta I_c$ spectroscopy measured empirically. The ratio of the magnitude of the central maximum for $B=0$ (corresponding to zero precession: $\theta=0$ and hence $\Delta\theta=0$) to the signal value for extrema at $\theta=n\pi$ ($n=1,2,3...$) is therefore

\begin{equation}
\frac{\left| \Delta I_c(0) \right|}{\left| \Delta I_c(\theta) \right|}=\exp{\frac{\Delta\theta^2}{2}}.
\end{equation}

Solving for $\Delta \theta$ gives 

\begin{equation}\label{eqn:dephasing0}
\Delta\theta \sim \sqrt{2\ln{\frac{\left| \Delta I_c(0)\right|}{\left| \Delta I_c(\theta)\right|}}}
\end{equation}

It is therefore possible to compare the dephasing effect in different experiments by studying the signal magnitude decrease of the same order extrema, without having to fit to the complete model (Eq. \ref{eqn:precession2}). Fig. \ref{fig:fig2}(b) shows the distribution width $\Delta\theta$ for the several precession oscillation extrema based on Eq. \ref{eqn:dephasing0} for simulations of devices with different transport layer thickness from $10 \mu m$ to $100 \mu m$. The same voltage bias, 2V, is applied on the transport layer. It is obvious that, despite an order-of-magnitude change in the transit length $L$, the dephasing is virtually constant, further confirming the conclusions drawn from Fig. \ref{fig:fig2}(a).

Voltage-controlled dephasing in a device with $L=350\mu m$ from our simulation is shown using Eq. \ref{eqn:dephasing0} in Fig. \ref{fig:fig3} (symbols). At fixed voltage, the changes in dephasing for successive extrema are approximately equal, because $\Delta\theta=g\mu_BB\Delta t/\hbar$ and the magnetic field period of precession oscillations is approximately constant. We compare the dephasing behavior of several precession oscillation extrema from these calculations to the behavior predicted by Eq. \ref{eqn:length} (solid lines). Despite there being no free parameters for fitting, we find excellent graphical agreement between Eqs. \ref{eqn:length} and \ref{eqn:dephasing0} for all precession oscillation extrema examined, confirming the validity of our obtained expressions.

\section{Experiments}

\begin{figure}
  \centering
  \includegraphics[scale=0.5]{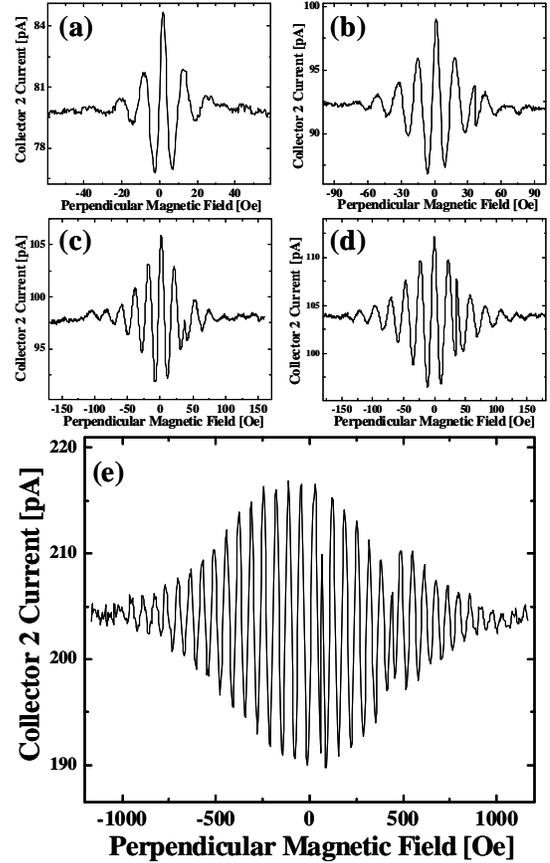}
  \caption{ \label{fig:fig5} (a)-(e): Experimentally-measured spin precession spectroscopy at different accelerating voltages 10V, 14V, 16V, 18V, and 80V, respectively, using spin transport through 350 micron-thick undoped single-crystal Si at 150K. Unprecedented spin coherence is evident in (e), showing at least 15 full spin rotations.}
\end{figure}

We now compare these results to experimental data, using all-electrical hot-electron spin injection and detection to study spin transport through 350-micron-thick undoped single-crystal Si(100)\cite{HUANG350}. The device scheme and corresponding band structure are shown in Fig. \ref{fig:fig4}(a) and (b). Transport proceeds from top to bottom: equilibrium spin-polarized electrons from the Co$_{84}$Fe$_{16}$ cathode tunnel through the Al$_2$O$_3$ barrier, becoming hot electrons in the nonmagnetic Al/Cu anode thin film. The electrons coupling with conduction band states over a Schottky barrier in the silicon transport layer on the other side forms the injected spin polarized current. Drift-dominated transport in the undoped silicon transport layer is controlled by $V_{C1}$, the ``accelerating voltage''. After transport through this silicon layer, the electrons are ejected from the conduction band and into the second ferromagnetic metal (Ni$_{80}$Fe$_{20}$) and the ballistic component of this current is collected by an n-type silicon substrate below, forming the spin-transport signal $I_{C2}$. This current is dependent on the projection of electron spin direction onto detector magnetization axis of the Ni$_{80}$Fe$_{20}$ layer\cite{APPELBAUMNATURE}. A magnetic field aligned perpendicular to the spin direction and parallel to internal electric field (and hence drift velocity) induces spin precession during transport from injector to detector. (This device is essentially the solid-state -- and Si-based -- version of the first bare-electron g-factor experiment which used spin precession during ballistic transport of high-energy electrons in vacuum and Mott scattering for spin filtering\cite{LOUISELL}.)

Fig. \ref{fig:fig5}(a)-(e) shows precession measurements at 150K using this device with accelerating voltage 10V, 14V, 16V, 18V and 80V, respectively. As the accelerating voltage increases, more oscillations are visible; at the highest voltage applied, more than 15 full spin rotations are evident by the number of oscillation extrema in Fig. \ref{fig:fig5}(e). The length independence derived earlier (Eq. \ref{eqn:length}) explains the high degree of spin coherence in these precession measurements, despite the 350 microns of Si separating injector from detector.

\begin{figure}
  \centering
  \includegraphics[scale=0.85]{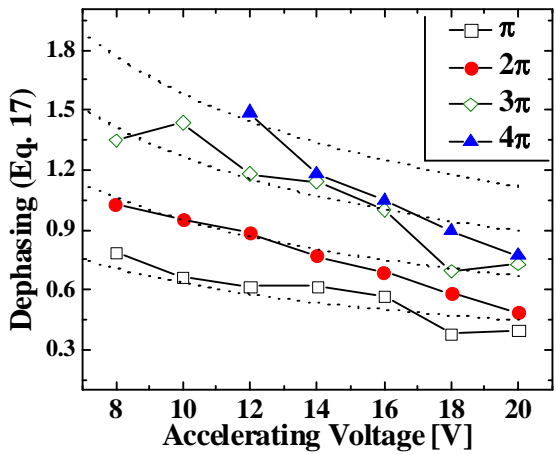}
  \caption{ \label{fig:fig6} Dephasing (calculated from experimental data with Eq. \ref{eqn:dephasing0}) for several precession oscillation extrema with different voltage bias across the 350-micron-thick undoped single-crystal Si transport layer at 150K. The model-predicted $V^{-1/2}$ behavior is also plotted for comparison (dotted line). }
\end{figure}

Using Eq. \ref{eqn:dephasing0}, we plot the dephasing at 150K as a function of voltage for several precession oscillation extrema in Fig. \ref{fig:fig6}, and compare to the $V^{-1/2}$ behavior predicted by Eq. \ref{eqn:length}.  Although the empirical behavior is qualitatively correct, the remaining discrepancy is likely due to our model assumption of ohmic transport, which at these voltages is not precisely upheld due to onset of velocity saturation\cite{BIQINJAP}. 

\begin{figure}
  \centering
  \includegraphics[scale=0.45]{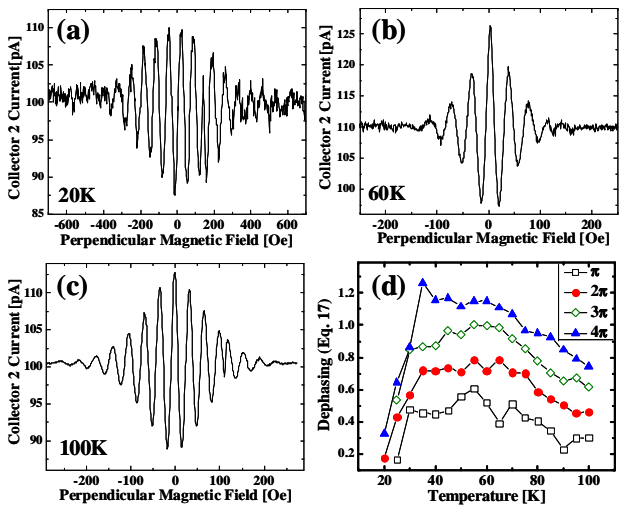}
  \caption{ \label{fig:fig7} (a)-(c) Temperature dependence on spin dephasing for our Si spin transport devices at three temperatures 20K, 60K, and 100K, respectively. Panel (d) plots the dephasing (calculated using Eq. \ref{eqn:dephasing0}) for 20K$<$T$<$100K, showing the unexpected non-monotonic dependence unaccounted for by application of the Einstein relation to our model. }
\end{figure}

The Einstein relation (or ``fluctuation-dissipation theorem''), valid for non-degenerate charge density, dictates $D/\mu=k_BT/q$, where $k_B$ is Boltzmann's constant and $T$ is temperature. Because our injected currents are low ($\approx 1-10\mu$A), and the spin-injector area is large ($\approx 50 \times 100 \mu$m$^2$), our experimental electron density $n=\frac{I\tau}{AqL}\approx 10^{11}$ cm${^3}$ is clearly non-degenerate in Si which has an effective conduction DOS $\approx 10^{19}$cm$^3$. If the Einstein relation is also valid for the non-equilibrium electrons we generate using ballistic electron injection with our tunnel-junction emitter, Eq. \ref{eqn:length} implies that the relative dephasing $\Delta\theta/\theta\propto \sqrt{T}$. Therefore, more dephasing should be evident at higher temperature.

Contrary to this expectation, we find a non-monotonic behavior of experimental dephasing as a function of temperature for experimental measurements at fixed accelerating voltage, as shown in Fig. \ref{fig:fig7}(a)-(c). The dephasing is small at the lowest temperature (20K), then rises to a maximum at approximately 60K, and drops gradually toward 100K, as shown in Fig. \ref{fig:fig7}(d). Clearly this observation is not compatible with the Einstein relation, and a more sophisticated model than the one presented here is required to capture the correct behavior. It is interesting to note, however, that the sharp decrease in dephasing at low temperatures corresponds coincidentally to the onset of a negative differential mobility (NDM) regime in Si(100) below $\approx$40K (Ref. \cite{JACOBONI}), which in other semiconductors gives rise to the Gunn effect: transport is via soliton-like spatial charge domains that suppress longitudinal diffusion and would therefore constrain spin dephasing in these spintronic devices. However, NDM occurs in Si(100) only in a window of electric field values ($\approx$50-150V/cm) that are substantially smaller than the ones used here. It is more likely that the observed reduction of dephasing is the result of the consequences of (unintentional) doping impurity freeze-out.

\section{Conclusion}
Using a spin precession model based on the Green's function of the spin drift-diffusion equation, we determined the expected effect of transit length and voltage changes on spin dephasing. Most importantly, we determined that for long spin lifetimes, the relative spin dephasing is independent of the transit length. In future spintronic devices, this length independency may enable multiple gate operations and long distance transport in precession-dependent devices as long as the finite spin lifetime allows. 

We also constructed an empirical measure of spin dephasing and used it to characterize both the results of simulation and experiment. Dephasing is shown to be inversely dependent on the square-root of voltage drop that drives drift, both theoretically and experimentally. 

For model results closer to experiment, the true 3-dimensional device geometry, magnetic fringe fields from injector and detector contacts, and non-ohmic transport, (all of which will give other systematic sources of dephasing) should be taken into account.

This work was supported by DARPA/MTO and ONR.

\end{document}